# Giant dielectric difference in chiral asymmetric bilayers

Yu-Hao Shen[1], Wen-Yi Tong[1], He Hu[1], Chun-Gang Duan[1,2*]

[1] *Key Laboratory of Polar Materials and Devices, Ministry of Education, East China Normal University, Shanghai, Shanghai 200241, China*

[2] *Collaborative Innovation Center of Extreme Optics, Shanxi University, Taiyuan, Shanxi 030006, China*

E-mail: cgduan@clpm.ecnu.edu.cn

**Abstract**

Twistronics rooted in the twist operation towards bilayer van der Waals crystals is of both theoretical and technological importance. The realize of the correlated electronic behaviors under this operation encourages enormous effort to the research on magic-angle systems which possess sensitive response to the external field. Here, a giant dielectric difference between 30± degree twist case is observed in a typical magnetic system 2H-VSe$_2$ bilayer. It is shown that due to the structural inversion asymmetry in its monolayer, the different stacking of the two cases corresponds to the two kind of valley polarized states: interlayer ferrovalley and interlayer antiferrovalley. Further investigations reveal that such different dielectric response between the two states stems from the different Fermi wave vectors coupled to the electric field. More interestingly, we even obtain the selective circularly polarized optical absorption by tuning the interlayer twist. These findings open an appealing route toward functional 2D materials design for electric and optical devices.

**Introduction**

Recent advances in fabrication of atomically thin materials have successfully realized the interlayer twist stacking of the bilayer two-dimensional (2D) van der Waals crystals, giving rise to an additional potential modulation[1-6]. In such 2D twisted systems, Moiré pattern with long period[7-13] can be introduced due to misoriented stacking. The patterned interlayer coupling in van der Waals crystals significantly modifies the low-energy band structure. As examples, the Moiré bands in twisted bilayer graphene (tBLG) and the Dirac velocity crossing zero several times with the change of the magic-angle[10, 14-16] have been reported. The strongly electron−electron correlation in tBLG yields various fascinating physical behaviors such as the transition from semimetal to Mott insulator[14, 16-18]. Unlike the magic-angle tBLG system with commensurate superlattice, when the twist angle is 30 degree between the two layers, the singularity appears, making the system lack of long period and incommensurate[19-22]. Any small relative rotation between layers will generate a finite supercell. It can be regarded as an infinite supercell possessing 12-fold rotational symmetry[21, 23], with combination of translational symmetry. The interlayer stacking of turning left (30+α degree twist) or right (30-α degree twist) case is equivalent for the bilayer graphene, which inspire us to break the chiral symmetry of the two commensurate partners to find more potential applications. Thanks to the staggering of two inequivalent sublattice (M=Mo, W and X=S, Se), two different stacking for the twisted transition-metal dichalcogenides (TMD) 2H-$MX_2$ system[24-36] can be generated. Experimentally, the information of interlayer stacking can be observed by acoustic measurement because the so-called Moiré excitons[31, 34] such as Moiré phonons[28] exhibit signatures of the Moiré patterned potential in this TMD-based system. Theoretically, model calculations[36, 37] have realized that the coupling between the ground states and excited states is tunable by interlayer twist. However, for most researches about Moiré pattern, the ferromagnetic TMD system has not been taken into account, which greatly restricts its potential applications in spintronics and valleytronics.

As a kind of typical 2D ferrovalley material[38, 39], the 2H-$VSe_2$ offers an ideal platform to explore the influence of interlayer twist on the valley polarization of its bilayer system. For our case of 30±α degree twist, we choose the interlayer magnetic coupling as antiferromagnetic coupling, which is the ground magnetic state of the pristine bilayer system[40, 41], and α as 2.2 degree. By performing first-principles calculations, we obtain totally different valley polarization for the whole bilayer system. It would be, as we defined, the interlayer ferrovalley state and interlayer antiferrovalley state for 30+α degree twist case and 30-α degree twist case, respectively. Furthermore, when applied the electric field vertical to the 2D plane, the interlayer ferrovalley state exhibits screening effect like metallic system, while the interlayer antiferrovalley state exhibits dielectric effect like insulating system, indicating a giant dielectric difference between the two states.

**Results**

We start by considering two relative rotation cases from 30-degree twisted system. i.e. 30±α degree twist of turning left or right, where we choose α as 2.2 degree in our case. And the interlayer antiferromagnetic order and intralayer ferromagnetic order are set for both cases. The top view of the incommensurate (infinite) and commensurate (finite) superlattices is shown in Fig.1(a). In fact, for infinite cell, all bands fold into its corresponding first Brillouin Zone (BZ), which is exactly the Gamma point. From the view of projection, the valley polarization of the two monolayers possess orthogonal, as shown in Fig.1(b). As for finite cell, where one layer rotates α degree with respect to the other, in its corresponding first BZ, the band folding leads to the parallel polarization for turning left case and antiparallel polarization for turning right case, respectively. Then we define the interlayer ferrovalley state and interlayer antiferrovalley state for 30±α degree twist case, respectively. Through calculating the spin polarized band structures with spin-orbit coupling (SOC), as shown in Fig.2(a), (g) for 27.8° twist case and Fig.2(b), (j) for 32.2° twist case, and analyzing the irreducible representations of K valley states, as illustrated in the left panel of Fig.4(a), (b), we confirm that the $K_+$ valley of the top layer is reversed to $K_-$ valley with respect to that of the bottom layer. This can be understood by the mirror symmetry of Moiré BZ of 30±α degree twist system along G-K axis[28], illustrated in dashed box Fig.1(b). The Moiré BZ is constructed from difference between the two $K_+$ (or $K_-$) wavevectors of the twisted first BZ and its basic vector $\vec{g} = \vec{G}_{top} - \vec{G}_{bottom}$[14, 28, 42, 43], where $\vec{G}_{top}$ and $\vec{G}_{bottom}$ is the reciprocal basic vector of the first BZ for two twisted monolayers. Since the Moiré superlattice of 30±α degree twist have the same lattice constant in a primitive cell, the $\vec{g}$ and $\vec{g}'$ of Moiré BZ for these two commensurate partners should satisfy: $|\vec{g}| = |\vec{g}'|$. As illustrated in the supplementary information of Ref[27], for the case of 30-α degree twist, the $\vec{g}$ is in the first BZ when we move g point to G point, whereas for the case of 30+α degree twist, the $\vec{g}$ is out of the first BZ and it has its equivalent vector $\vec{g}'$ which we can find in the translational first BZ of one layer with a $\vec{G}$ vector. Then for the Moiré BZ with $\vec{g}'$ vector, it is equivalently constructed from the difference between the K+ (K-) wavevector of the first BZ for one layer and K- (K+) wavevector of that for the other one. Hence, for 30±α degree twisted system, we obtain two different superposition of band structures for two layers. However, macroscopically, it is impossible to distinguish the two stacking cases by optical means since the K valley states for exciting left-handed and right-handed circularly polarized light is degenerate energetically in both cases. An applied vertical bias voltage leading to a potential difference between layers can break the energy degeneracy[41, 44], where the sub-bands from one layer move upward with respect to those from the other, reducing the energy gap of the whole system. And the interlayer coupling is modified under vertical electric field[45-47]. Following this strategy, we then apply external electric field vertical to the 2D plane, pointing from top layer to bottom layer and find that, as shown in Fig.2, previous energy degeneracy between two layers is decoupled with an interlayer potential difference. The electric field moves the sub-bands of the bottom layer downward and that of the top layer upward. However, the reduction of the energy gap for 27.8° twist system (shown in Fig.2(c), (e), (h), (i)) is much larger than that for 32.2° twist system (shown in Fig.2(d), (f), (k), (l)). This different dielectric response can be attributed to the different Fermi wave vectors coupled to the electric field, leading to the significantly different charge screening of the two twisted system, shown as the induced charge densities Δρ = ρ(E=0.02 V/Å) − ρ(E=0.00 V/Å) in Fig.4(c), where the screening charges distribute differently between layers (shown in the dashed box). It is indicated that the 30±α degree twist of turning left or right system possess significantly different dielectric constant. Under enough large electric field vertical to the 2D plane, for turning right case, the valence bands near K valley of one layer is lifted above Fermi level while that of the other layer is still below Fermi level and for turning left case, all bands are below Fermi level but the bands of one layer is always above that of the other layer. Hence, in considering of the spin valley coupling, with respect to 30-degree twisted stacking case, only the left-handed circularly polarized light can be absorbed by the bottom layer of turning right case (right panel of Fig.4(a)) and only the right-handed circularly polarized light can be absorbed by the top layer of turning left case (right panel of Fig.4(b)). Then we obtain the selective circularly polarized optical absorption by tuning interlayer twist. Furthermore, without SOC effect, we choose not too large α as 2.2°, 8.2° and 12.1°, and then calculate the spin polarized density of states (DOS) for these 30±α degree twist case under a value of 0.02 V/Å electric field. As shown in Fig.3, for both 30±α degree twist case, the energy gap of both spin up and spin down sub-bands become larger as the α increases and its change is far greater than the change of the electric induced energy shift δ between two spin sub-bands, indicating the giant dielectric difference is almost independent of the specific α.

**Discussion**

To explore the mechanism of giant dielectric difference in 30±α degree twisted system, we develop the two-band effective Hamiltonian of the twisted bilayer[48]:

$$H(\vec{k}) = \begin{pmatrix} \epsilon_{\vec{k}}^t & 0 \\ 0 & \epsilon_{\vec{k}}^b \end{pmatrix} + \begin{pmatrix} \langle \phi_{\vec{k}}^t | \vec{V}_b | \phi_{\vec{k}}^t \rangle & \langle \phi_{\vec{k}}^t | T + \vec{V}_t + \vec{V}_b | \phi_{\vec{k}}^b \rangle \\ \langle \phi_{\vec{k}}^b | T + \vec{V}_t + \vec{V}_b | \phi_{\vec{k}}^t \rangle & \langle \phi_{\vec{k}}^b | \vec{V}_t | \phi_{\vec{k}}^b \rangle \end{pmatrix}$$

where $b$ and $t$ refer to bottom and top monolayer, respectively. The kinetic term is $T = \frac{p^2}{2m}$ and the potential term $\vec{V}_t$ and $\vec{V}_b$ have monolayer translation symmetry, respectively. A complete basis for the bilayer is that formed from the eigen vectors $\{\phi_{\vec{k}}^t, \phi_{\vec{k}}^b\}$
of the monolayer, which satisfy:

$$(T + \vec{V}_t)\phi_{\vec{k}}^t = \epsilon_{\vec{k}}^t \phi_{\vec{k}}^t$$

$$(T + \vec{V}_b)\phi_{\vec{k}}^b = \epsilon_{\vec{k}}^b \phi_{\vec{k}}^b$$

Note that for bilayer 2H-VSe$_2$, its valence band maximum (VBM) is located in the Gamma point. We then introduce k·p perturbation into the effective Hamiltonian and concentrate on the energy shift of the band reference to the Fermi level. In consideration of the intralayer ferromagnetic coupling and interlayer antiferromagnetic coupling, for spin down (up) states, $\epsilon_{\vec{k}}^t$ and $\epsilon_{\vec{k}}^b$ is the corresponding energy of the electron (hole) band and hole (electron) band, respectively. In addition, as shown in the Fig.5, the twisted first Brillouin Zone (BZ) of the two layers produces a momentum shift $\vec{q} = \vec{k}_t - \vec{k}_b$ between electron and hole bands in the Moiré BZ, where $\vec{k}_t$ and $\vec{k}_b$ is the Bloch wave vector for the top and bottom layer, respectively. In this way, the $\vec{V}_t$ and $\vec{V}_b$ can be expressed as staggered Moiré potential form:

$$\frac{p^2}{2m} + \vec{V}_t = -\frac{\hbar^2 q^2}{2m^*} + sV(\vec{r})$$

$$\frac{p^2}{2m} + \vec{V}_b = -\frac{\hbar^2 q^2}{2m^*} - sV(\vec{r})$$

where the $s = \pm 1$ is the spin index, $m^*$ is the valence band effective mass and $q$ is the momentum measured from g point in the Moiré BZ. Since the Bloch wave vector $\vec{k}_t$ and $\vec{k}_b$ are coupled only when $\vec{k}_t + \vec{G}_{top} = \vec{k}_b + \vec{G}_{bottom}$, where $\vec{G}_{top}$ and $\vec{G}_{bottom}$ is the reciprocal basic vector of the first BZ for two twisted layers[49, 50], the Moiré potential[35] $V(\vec{r})$ can be Fourier transformed with basic vector $\vec{g} = \vec{G}_{top} - \vec{G}_{bottom}$ as: $V(\vec{r}) = \sum'_{\vec{g}} V(\vec{g}) e^{i\vec{g}*\vec{r}}$. We introduce the layer index 1 and 2 for top and bottom monolayer, respectively. Then the k·p effective Hamiltonian near g point can be expressed as:

$$H = \begin{pmatrix} \epsilon_1 + \frac{\Delta_\alpha}{2} - \langle \phi_1 | sV | \phi_1 \rangle & \langle \phi_1 | T | \phi_2 \rangle \\ \langle \phi_2 | T | \phi_1 \rangle & \epsilon_2 - \frac{\Delta_\alpha}{2} + \langle \phi_2 | sV | \phi_2 \rangle \end{pmatrix}$$

$\epsilon_1$, $\epsilon_2$ and $\phi_1$, $\phi_2$ satisfy:

$$(T + sV)\phi_1 = \epsilon_1 \phi_1$$

for the top monolayer

$$(T - sV)\phi_2 = \epsilon_2 \phi_2$$

for the bottom monolayer

where the $T = -\frac{\hbar^2 q^2}{2m^*}$ is the kinetic energy term and $\Delta_\alpha$ is the gap correlation depending on the specific $\alpha$. For 30-degree twisted system, $\alpha = 0$, the constructed Moiré BZ is illustrated in Fig.5(a). As the twist angle become $30\pm\alpha$ degree, $\alpha > 0$ and it is not too large, then the modified Fermi wave vector of the whole system can be illustrated as the g point shift to g' point, shown in Fig.4(b) of 30+α degree twist case, which is $\Delta \vec{k}_F = \Delta \vec{k}_F^1 + \Delta \vec{k}_F^2$, where $\Delta \vec{k}_F^1$ and $\Delta \vec{k}_F^2$ corresponds to the contribution of top and bottom layer, respectively and $|\Delta \vec{k}_F^1| = |\Delta \vec{k}_F^2|$. In fact, they stem from the modification $\delta V$ towards the Moiré potential $V$, where $\delta V \ll V$. Then the effective Hamiltonian can be expressed as:

$$H = \begin{pmatrix} \epsilon_1 + \frac{\Delta_\alpha}{2} + \langle \phi_1 | -sV + s\delta V | \phi_1 \rangle & \langle \phi_1 | T | \phi_2 \rangle \\ \langle \phi_2 | T | \phi_1 \rangle & \epsilon_2 - \frac{\Delta_\alpha}{2} + \langle \phi_2 | sV - s\delta V | \phi_2 \rangle \end{pmatrix}$$

$$H = \begin{pmatrix} \epsilon_1 + \frac{\Delta_\alpha}{2} + \langle\phi_1|-sV - s\delta V|\phi_1\rangle & \langle\phi_1|T|\phi_2\rangle \\ \langle\phi_2|T|\phi_1\rangle & \epsilon_2 - \frac{\Delta_\alpha}{2} + \langle\phi_2|sV + s\delta V|\phi_2\rangle \end{pmatrix}$$

for 30+α degree twist case.

Under applied electric field to the 2D plane, pointing from top layer to bottom layer, the spatial symmetry of D$_3$ point group is broken. To clarify this, we consider the momentum shift of reference point (G to G') between layers due to the interlayer potential difference, as shown in Fig.5(c) for 30-degree twist case, so as g shifting to g' of corresponding Moiré BZ. The broken symmetry leads to different modification $\delta V$ (or $\delta V'$) towards the Moiré potential $V$, illustrated as $|\Delta \vec{k}_F^1| > |\Delta \vec{k}_F^2|$, shown in Fig.4(d) of 30+α degree twist case. Here we assume $\delta V > \delta V'$ in our case. Then the effective Hamiltonian can be expressed as:

$$H = \begin{pmatrix} \epsilon_1 + \frac{\Delta_\alpha}{2} + \langle\phi_1|-sV + s\delta V|\phi_1\rangle & \langle\phi_1|T + s\delta V - s\delta V'|\phi_2\rangle \\ \langle\phi_2|T + s\delta V - s\delta V'|\phi_1\rangle & \epsilon_2 - \frac{\Delta_\alpha}{2} + \langle\phi_2|sV - s\delta V'|\phi_2\rangle \end{pmatrix}$$

for 30-α degree twist case.

$$H = \begin{pmatrix} \epsilon_1 + \frac{\Delta_\alpha}{2} + \langle\phi_1|-sV - s\delta V|\phi_1\rangle & \langle\phi_1|T - s\delta V + s\delta V'|\phi_2\rangle \\ \langle\phi_2|T - s\delta V + s\delta V'|\phi_1\rangle & \epsilon_2 - \frac{\Delta_\alpha}{2} + \langle\phi_2|sV + s\delta V'|\phi_2\rangle \end{pmatrix}$$

for 30+α degree twist case.

And the $\epsilon_1$, $\epsilon_2$ and $\phi_1$, $\phi_2$ satisfy:

$$(T + sV + H_E)\phi_1 = \epsilon_1 \phi_1$$

for the top monolayer

$$(T - sV - H_E)\phi_2 = \epsilon_2 \phi_2$$

for the bottom monolayer

Here the $H_E$ is the electric potential term. When $s = -1$, the eigen energy values of the two band (the electron band and the hole band) for 30-degree twist case can be expressed as:

$$E_{\pm} = \frac{1}{2}(\epsilon_1 + \epsilon_2 + \langle\phi_1|V|\phi_1\rangle - \langle\phi_2|V|\phi_2\rangle \pm \sqrt{E_0})$$

Where

$$E_0 = (\epsilon_1 - \epsilon_2 + \Delta_0)^2 + 4\langle\phi_1|T|\phi_2\rangle\langle\phi_2|T|\phi_1\rangle + 2(\epsilon_1 - \epsilon_2 + \Delta_0)(\langle\phi_1|V|\phi_1\rangle + \langle\phi_2|V|\phi_2\rangle)$$

$$+ (\langle\phi_1|V|\phi_1\rangle + \langle\phi_2|V|\phi_2\rangle)^2$$

The change of the energy gap for the $s = -1$ sub-bands can be reflected by the change of the energy distance $\Delta$ between the two eigen energy values, which is:

$$\Delta = \sqrt{E_0}$$

for 30-degree twist case.

$$\Delta = \sqrt{E_\alpha - \Delta_T - \Delta_E - \Delta'_\alpha}$$

for 30-α degree twist case.

$$\Delta = \sqrt{E_\alpha + \Delta_T + \Delta_E + \Delta'_\alpha}$$

for 30+α degree twist case.

Where

$$E_\alpha = (\epsilon_1 - \epsilon_2 + \Delta_\alpha)^2 + 4\langle\phi_1|T|\phi_2\rangle\langle\phi_2|T|\phi_1\rangle + 2(\epsilon_1 - \epsilon_2 + \Delta_\alpha)(\langle\phi_1|V|\phi_1\rangle + \langle\phi_2|V|\phi_2\rangle)$$

$$+ (\langle\phi_1|V|\phi_1\rangle + \langle\phi_2|V|\phi_2\rangle)^2$$

and

$$\Delta_T = 4\langle\phi_1|\delta V - \delta V'|\phi_2\rangle\langle\phi_2|T|\phi_1\rangle + 4\langle\phi_2|\delta V - \delta V'|\phi_1\rangle\langle\phi_1|T|\phi_2\rangle$$

is the twist contribution from interlayer electron tunneling.

$$\Delta_E = 2(\langle\phi_1|\delta V|\phi_1\rangle + \langle\phi_2|\delta V'|\phi_2\rangle)[\langle\phi_1|T + H_E|\phi_1\rangle - \langle\phi_2|T - H_E|\phi_2\rangle]$$

is the twist contribution from intralayer electron hopping under interlayer potential difference.

$$\Delta'_\alpha = \Delta_\alpha(\langle\phi_1|\delta V|\phi_1\rangle + \langle\phi_2|\delta V'|\phi_2\rangle)$$

is the $\alpha$ dependent twist contribution.

Here we neglect the quadratic term related to $\delta V$ and $\delta V'$. And the neglected change of the δ (shown in Fig.3) indicates $\Delta'_\alpha \ll \Delta_E$, which can also be neglected. Since both the $\Delta_T$ and $\Delta_E$ possess positive contribution, reference to the $\Delta = $ $\sqrt{E_\alpha}$ the $\Delta$ of $s=-1$ sub-bands decrease for 30-α degree twist case and increase for 30+α degree twist case, respectively. For the case of $s=+1$ sub-bands, it is demonstrated that the corresponding $\Delta$ possess opposite trend, compared to the case of $s=-1$ sub-bands. The change of the $\Delta$ is consistent with the change of the energy gap from DFT calculations:

$$\Delta_{g(30-\alpha)}^{\downarrow} < \Delta_{g(30+\alpha)}^{\downarrow}$$

$$\Delta_{g(30-\alpha)}^{\uparrow} > \Delta_{g(30+\alpha)}^{\uparrow}$$

Here, $\Delta_g^{\uparrow}$ and $\Delta_g^{\downarrow}$ represents the energy gap of the spin up and spin down sub-bands, respectively. Based on above qualitive discussion, it is indicated that the dielectric difference in 30±α degree twisted system can be attributed to the different Fermi wave vectors coupled to the electric field. It will enhance or reduce the electric screening effect of the twisted system.

To summarize, our theoretical investigations reveal that, due to the chiral asymmetry between 30± degree twisted bilayer of 2H-VSe$_2$, the different stackings leads to the two defined valley polarized states i.e. interlayer ferrovalley and antiferrovalley. When a vertical electric field is applied, the valley degeneracy is broken. The giant dielectric difference of the two stacking cases appears. We then construct a k·p model Hamiltonian to qualitatively analyze the mechanism towards the significantly different change of the energy gap. More importantly, the potential applications for utilizing this difference is, near 30 degree, the selective circularly polarized optical absorption can be obtained by tuning the interlayer twist with a tiny angle. Under large enough electric field, right-handed or left-handed light can be selectively absorbed when we turn left or right one layer with respect to the other layer. It would be exciting to realize the giant dielectric difference produced by chiral asymmetric bilayer system, which will be of great practical significance in the 2D spintronic and valleytronic devices.

**Method**

The calculations of bilayer VSe$_2$ are performed within DFT using the projector augmented wave (PAW) method implemented in the Vienna ab initio Simulation Package (VASP)[51]. The exchange-correlation potential is treated in Perdew–Burke–Ernzerhof form[52] of the generalized gradient approximation (GGA) with a kinetic-energy cutoff of 400 eV. A well-converged 9×9×1 Monkhorst-Pack k-point mesh is chosen in self-consistent calculations. The convergence criterion for the electronic energy is $10^{-5}$ eV and the structures are relaxed until the Hellmann–Feynman forces on each atom are less than 1 meV/Å. In our calculations, dispersion corrected DFT-D2 method[53] is adopted to describe the van der Waals interactions. The external electric field is introduced by planar dipole layer method.

**Acknowledgements**

This work was supported by the National Key Project for Basic Research of China (Grant Nos. 2014CB921104, SQ2017YFJC010044-03), the National Natural Science Foundation of China (Grant Nos. 11774092, 51572085, 61774059) and the Shanghai Science and Technology Innovation Action Plan (No. 17JC1402500). Computations were performed at the ECNU computing center.

**Figures**

FIG 1. (a) Top view of the bilayer 2H-VSe$_2$ commensurate cell of 27.8° twist case (right panel), 32.2° twist case (left panel) and incommensurate cell of 30° twist case (middle panel). Grass-green, red, and blue spheres represent selenium, spin-up, and spin-down vanadium atoms, respectively. (b) Schematic views of the valley polarization of the two individual layers in twisted system. In considering of band folding, the valley polarization in the first Brillouin Zone (BZ) of the commensurate and incommensurate supercell is illustrated schematically in the dashed box. The Moiré BZ is constructed from difference between the two K$_+$ (or K$_-$) wavevectors for the two layers, which is generally used in model calculations of twisted bilayer system.

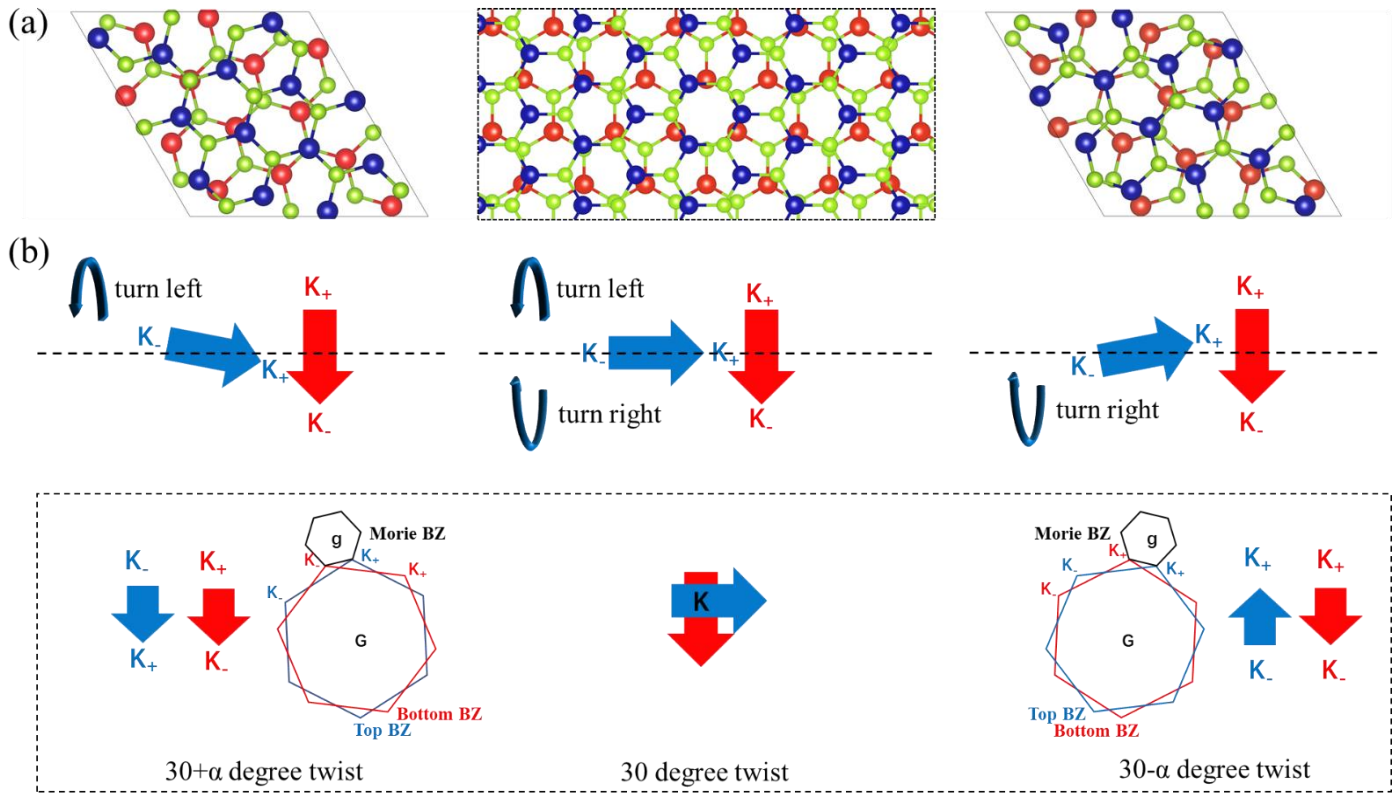

FIG 2. Orbital- and spin-resolved band structures with spin orbit coupling (SOC) of (a-f) $V - d_{xy}$ projection and (g-l) $V - d_{z^2}$ projection for (a, c, e and g, h, i) 27.8° twist case and (b, d, f and j, k, l) 32.2° twist case. With electric field applied, all sub-bands of top layer are lifted up while sub-bands of bottom layer are pushed down.

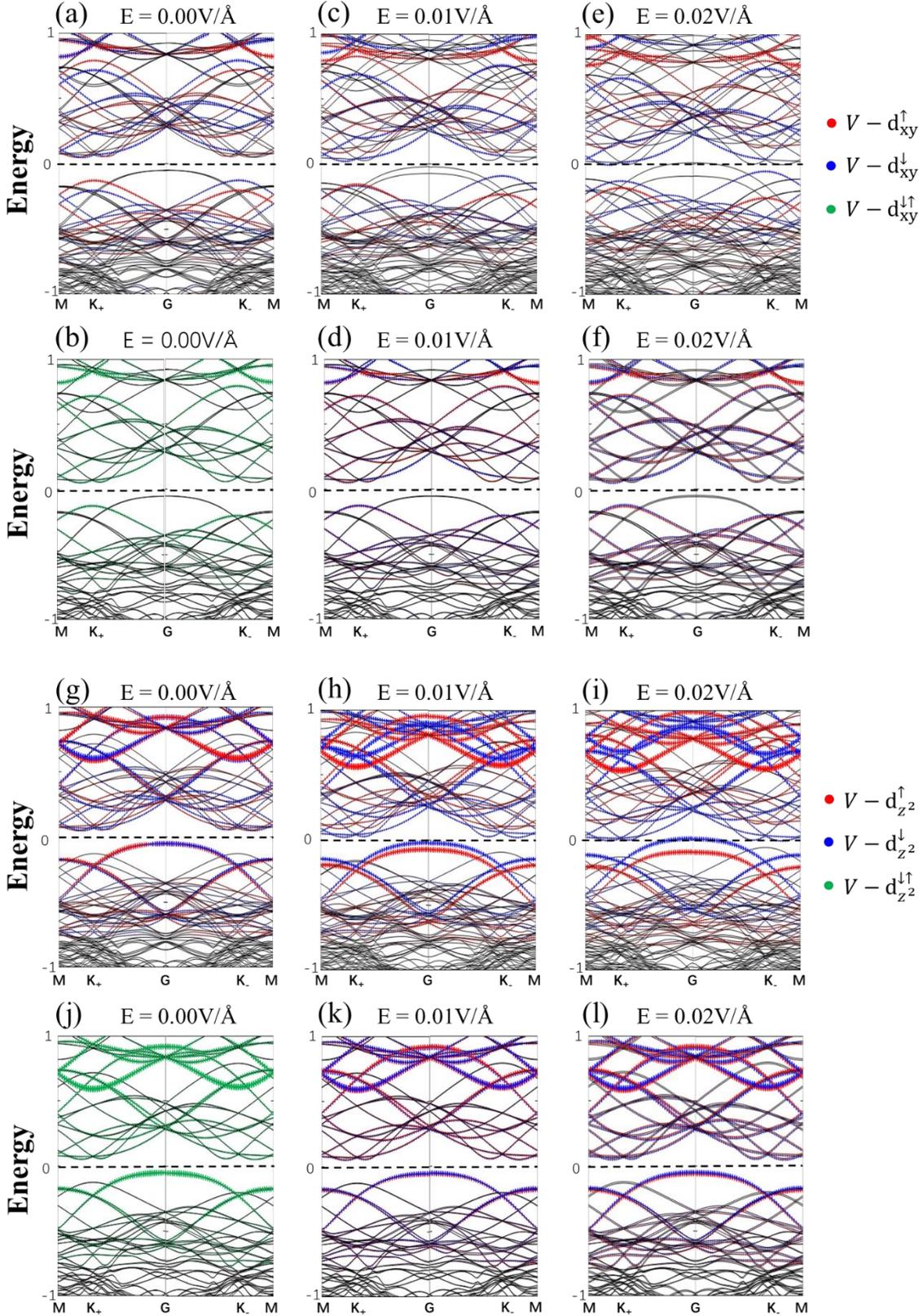

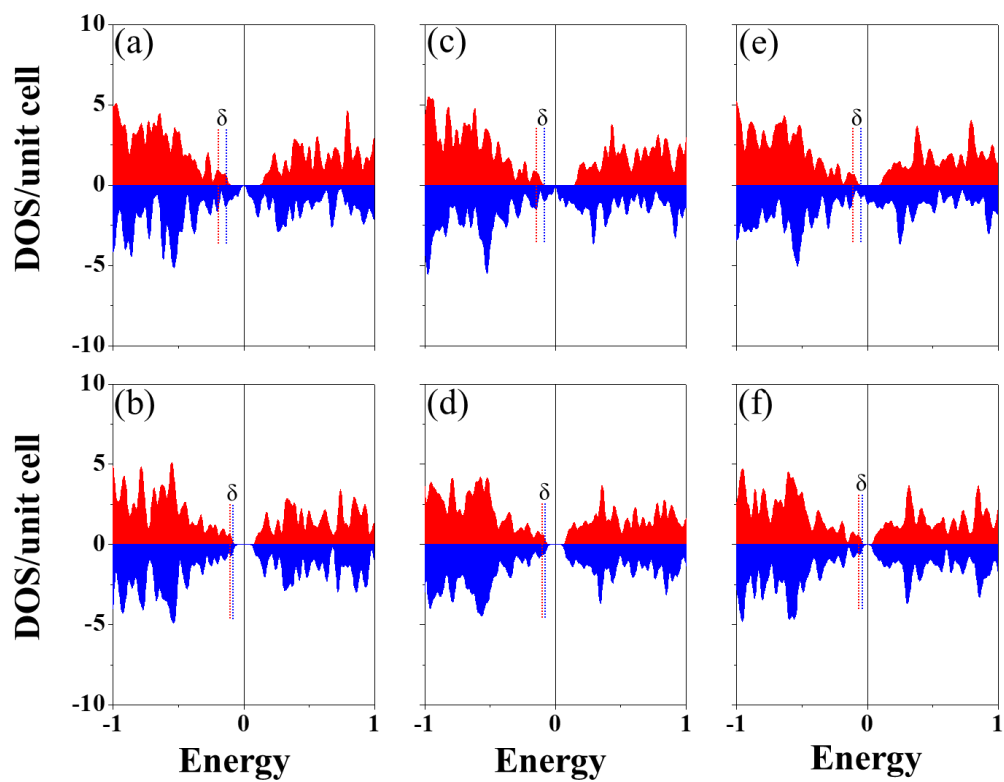

FIG 3. The calculated spin polarized density of states (DOS) without spin orbit coupling (SOC) under a value of 0.02 V/Å electric field. (a) 17.9° twist case. (b) 42.1° twist case. (c) 21.7° twist case. (d) 38.2° twist case. (e) 27.8° twist case. (f) 32.2° twist case. The δ denotes the electric induced energy shift between two spin sub-bands. Red area and blue area represent the spin up and spin down channel, respectively.

FIG 4. Schematic band structures near K± based on the calculated results in Fig.2 for (a) 32.2° twist case and (b) 27.8° twist case without electric field (left panel) and with electric field (right panel). Optical selection rules for the top valence band are plotted. $\sigma_+$ and $\sigma_-$ represents left-handed and right-handed circularly polarized optical adsorption, respectively. The irreducible representations of states have been labeled using the Mülliken notations. Dashed and solid curved line represents the K valley states from top layer and bottom layer, respectively. Green curved line represents the coincide of spin-up K valley states (red) and spin-down K valley states (blue). (c) The electric field-induced charge densities $\Delta\rho = \rho(E) - \rho(0)$ in arbitrary units under the influence of the electric field E = 0.02 V/Å for 27.8° twist case (left panel) and 32.2° twist case (right panel). Yellow and cyan colors represent the accumulation and depletion of electrons, respectively.

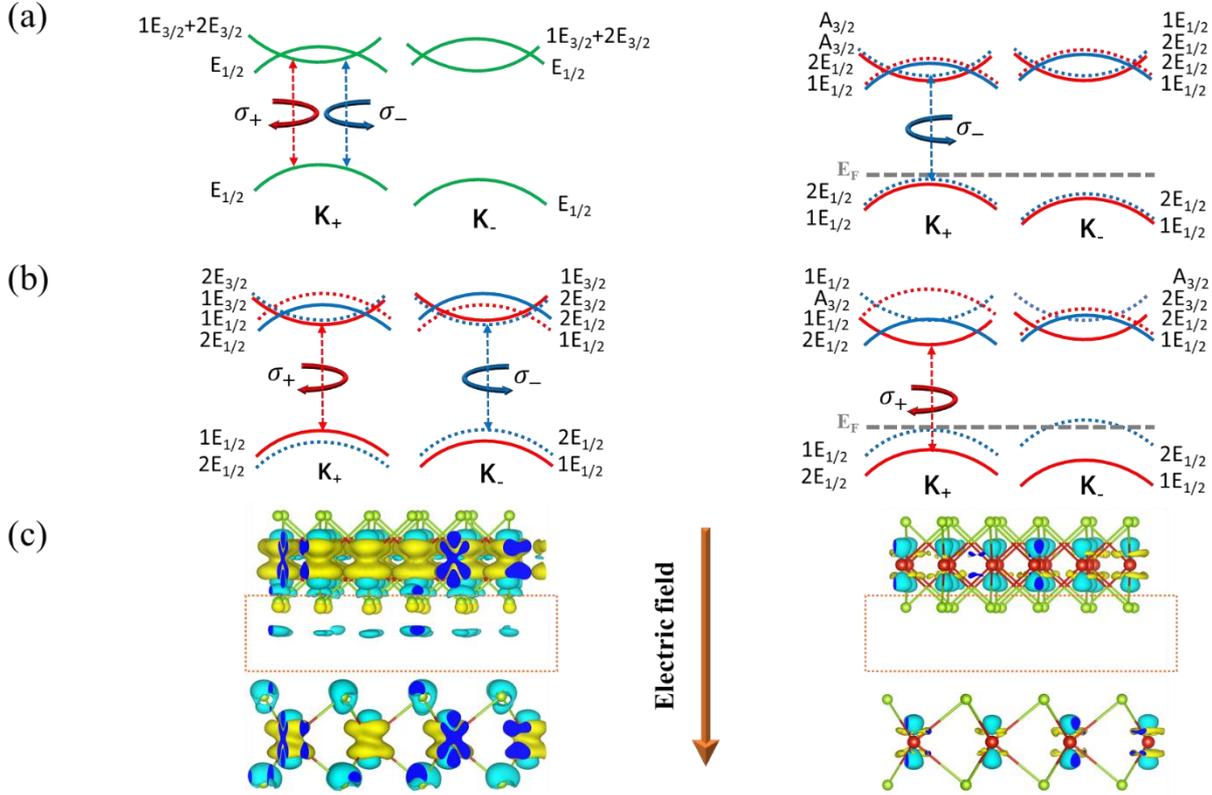

FIG 5. Scheme for analyzing the modified Fermi wave vector $\Delta \vec{k}_F = \Delta \vec{k}_F^1 + \Delta \vec{k}_F^2$ by interlayer twist from 30 degree in constructed Moiré BZ, where $\Delta \vec{k}_F^1$ and $\Delta \vec{k}_F^2$ corresponds to the contribution of top and bottom layer, respectively. (a) 30-degree twist case. (b) 30+α degree twist case. (c) 30-degree twist case with interlayer potential difference. (d) 30+α degree twist case with interlayer potential difference. The black orange and purple hexagon denote the Moiré BZ of 30-degree twist case, the Moiré BZ of 30+α degree twist case and the Moiré BZ of 30-degree twist case with interlayer potential difference, respectively.

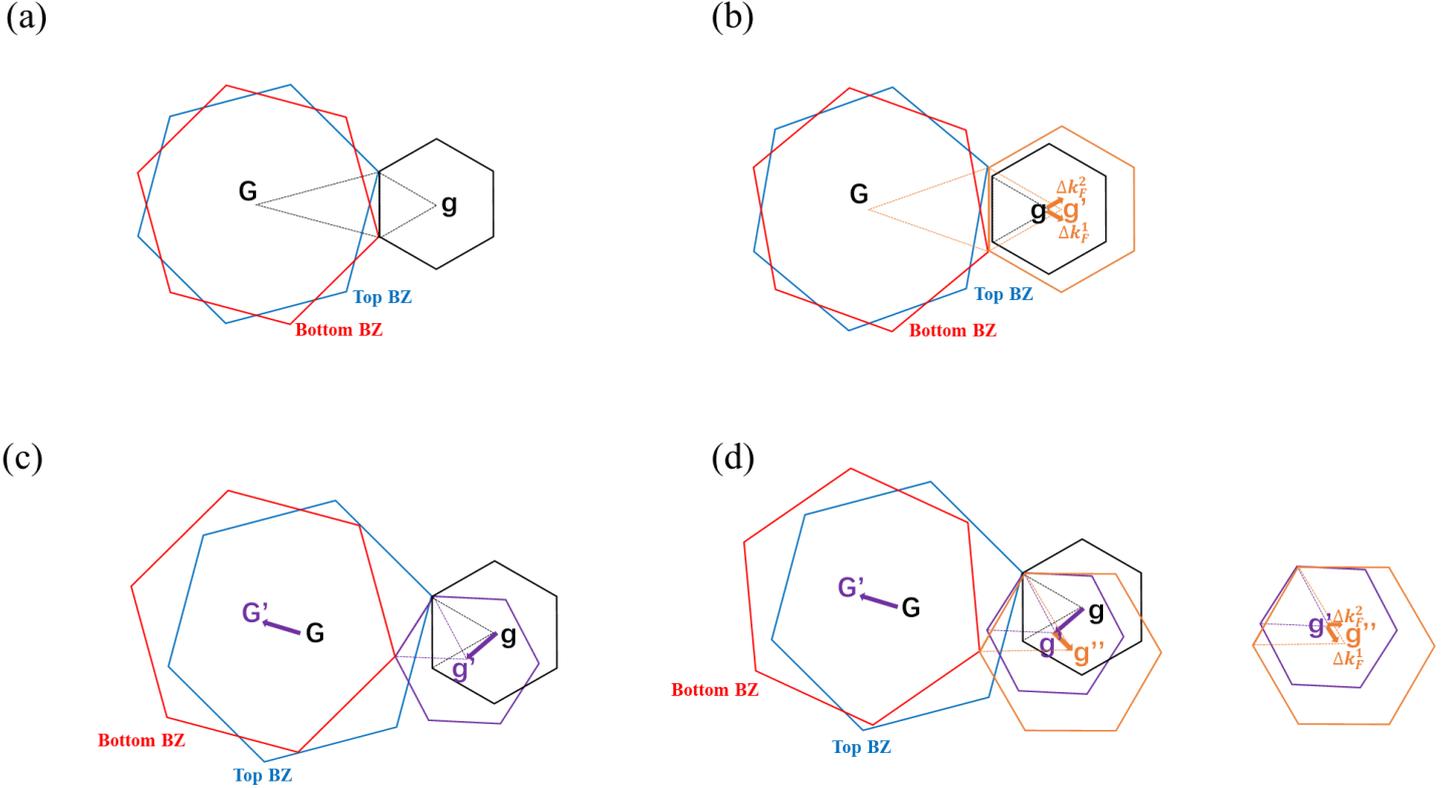

## References

1. Berger, C.; Song, Z. M.; Li, X. B.; Wu, X. S.; Brown, N.; Naud, C.; Mayou, D.; Li, T. B.; Hass, J.; Marchenkov, A. N.; Conrad, E. H.; First, P. N.; de Heer, W. A. *Science* **2006,** 312, (5777), 1191-1196.
2. Ni, Z. H.; Wang, Y. Y.; Yu, T.; You, Y. M.; Shen, Z. X. *Physical Review B* **2008,** 77, (23), 235403.
3. Yan, Z.; Peng, Z. W.; Sun, Z. Z.; Yao, J.; Zhu, Y.; Liu, Z.; Ajayan, P. M.; Tour, J. M. *Acs Nano* **2011,** 5, (10), 8187-8192.
4. Xie, L. M.; Wang, H. L.; Jin, C. H.; Wang, X. R.; Jiao, L. Y.; Suenaga, K.; Dai, H. J. *Journal of the American Chemical Society* **2011,** 133, (27), 10394-10397.
5. Zhao, R. Q.; Zhang, Y. F.; Gao, T.; Gao, Y. B.; Liu, N.; Fu, L.; Liu, Z. F. *Nano Research* **2011,** 4, (7), 712-721.
6. Carr, S.; Massatt, D.; Fang, S.; Cazeaux, P.; Luskin, M.; Kaxiras, E. *Physical Review B* **2017,** 95, (7), 075420.
7. Kang, J.; Li, J.; Li, S. S.; Xia, J. B.; Wang, L. W. *Nano Letters* **2013,** 13, (11), 5485-5490.
8. Li, G. H.; Luican, A.; dos Santos, J. M. B. L.; Castro Neto, A. H.; Reina, A.; Kong, J.; Andrei, E. Y. *Nature Physics* **2010,** 6, (2), 109-113.
9. Ugeda, M. M.; Bradley, A. J.; Shi, S. F.; da Jornada, F. H.; Zhang, Y.; Qiu, D. Y.; Ruan, W.; Mo, S. K.; Hussain, Z.; Shen, Z. X.; Wang, F.; Louie, S. G.; Crommie, M. F. *Nature Materials* **2014,** 13, (12), 1091-1095.
10. Bistritzer, R.; MacDonald, A. H. *Proceedings of the National Academy of Sciences of the United States of America* **2011,** 108, (30), 12233-12237.
11. Yankowitz, M.; Xue, J. M.; Cormode, D.; Sanchez-Yamagishi, J. D.; Watanabe, K.; Taniguchi, T.; Jarillo-Herrero, P.; Jacquod, P.; LeRoy, B. J. *Nature Physics* **2012,** 8, (5), 382-386.
12. Xue, J. M.; Sanchez-Yamagishi, J.; Bulmash, D.; Jacquod, P.; Deshpande, A.; Watanabe, K.; Taniguchi, T.; Jarillo-Herrero, P.; Leroy, B. J. *Nature Materials* **2011,** 10, (4), 282-285.

13. Luican, A.; Li, G. H.; Reina, A.; Kong, J.; Nair, R. R.; Novoselov, K. S.; Geim, A. K.; Andrei, E. Y. *Physical Review Letters* **2011,** 106, (12), 126802.
14. Cao, Y.; Fatemi, V.; Demir, A.; Fang, S.; Tomarken, S. L.; Luo, J. Y.; Sanchez-Yamagishi, J. D.; Watanabe, K.; Taniguchi, T.; Kaxiras, E.; Ashoori, R. C.; Jarillo-Herrero, P. *Nature* **2018,** 556, (7699), 80.
15. de Laissardiere, G. T.; Mayou, D.; Magaud, L. *Physical Review B* **2012,** 86, (12), 125413.
16. Padhi, B.; Setty, C.; Phillips, P. W. *Nano Letters* **2018,** 18, (10), 6175-6180.
17. Kim, K.; DaSilva, A.; Huang, S.; Fallahazad, B.; Larentis, S.; Taniguchi, T.; Watanabe, K.; Leroy, B. J.; MacDonald, A. H.; Tutuc, E. *Proceedings of the National Academy of Sciences of the United States of America* **2017,** 114, (13), 3364-3369.
18. Cao, Y.; Fatemi, V.; Fang, S.; Watanabe, K.; Taniguchi, T.; Kaxiras, E.; Jarillo-Herrero, P. *Nature* **2018,** 556, (7699), 43.
19. Woods, C. R.; Britnell, L.; Eckmann, A.; Ma, R. S.; Lu, J. C.; Guo, H. M.; Lin, X.; Yu, G. L.; Cao, Y.; Gorbachev, R. V.; Kretinin, A. V.; Park, J.; Ponomarenko, L. A.; Katsnelson, M. I.; Gornostyrev, Y. N.; Watanabe, K.; Taniguchi, T.; Casiraghi, C.; Gao, H. J.; Geim, A. K.; Novoselov, K. S. *Nature Physics* **2014,** 10, (6), 451-456.
20. Lebedev, A. V.; Lebedeva, I. V.; Popov, A. M.; Knizhnik, A. A. *Physical Review B* **2017,** 96, (8), 085432.
21. Yao, W.; Wang, E. Y.; Bao, C. H.; Zhang, Y. O.; Zhang, K. A.; Bao, K. J.; Chan, C. K.; Chen, C. Y.; Avila, J.; Asensio, M. C.; Zhu, J. Y.; Zhou, S. Y. *Proceedings of the National Academy of Sciences of the United States of America* **2018,** 115, (27), 6928-6933.
22. Lebedeva, I. V.; Lebedev, A. V.; Popov, A. M.; Knizhnik, A. A. *Physical Review B* **2016,** 93, (23), 235414.
23. Ahn, S. J.; Moon, P.; Kim, T. H.; Kim, H. W.; Shin, H. C.; Kim, E. H.; Cha, H. W.; Kahng, S. J.; Kim, P.; Koshino, M.; Son, Y. W.; Yang, C. W.; Ahn, J. R. *Science* **2018,** 361, (6404), 782.
24. Wang, K.; Huang, B.; Tian, M.; Ceballos, F.; Lin, M. W.; Mahjouri-Samani, M.; Boulesbaa, A.; Puretzky, A. A.; Rouleau, C. M.; Yoon, M.; Zhao, H.; Xiao, K.; Duscher, G.; Geohegan, D. B. *ACS Nano* **2016,** 10, (7), 6612-6622.
25. Zheng, S. J.; Sun, L. F.; Zhou, X. H.; Liu, F. C.; Liu, Z.; Shen, Z. X.; Fan, H. J. *Advanced Optical Materials* **2015,** 3, (11), 1600-1605.
26. Liu, K. H.; Zhang, L. M.; Cao, T.; Jin, C. H.; Qiu, D. A.; Zhou, Q.; Zettl, A.; Yang, P. D.; Louie, S. G.; Wang, F. *Nature Communications* **2014,** 5, 4966.
27. Huang, S. X.; Liang, L. B.; Ling, X.; Puretzky, A. A.; Geohegan, D. B.; Sumpter, B. G.; Kong, J.; Meunier, V.; Dresselhaus, M. S. *Nano Letters* **2016,** 16, (2), 1435-1444.
28. Lin, M. L.; Tan, Q. H.; Wu, J. B.; Chen, X. S.; Wang, J. H.; Pan, Y. H.; Zhang, X.; Cong, X.; Zhang, J.; Ji, W.; Hu, P. A.; Liu, K. H.; Tan, P. H. *Acs Nano* **2018,** 12, (8), 8770-8780.
29. Wang, Z. L.; Chen, Q.; Wang, J. L. *Journal of Physical Chemistry C* **2015,** 119, (9), 4752-4758.
30. Lu, N.; Guo, H. Y.; Zhuo, Z. W.; Wang, L.; Wu, X. J.; Zeng, X. C. *Nanoscale* **2017,** 9, (48), 19131-19138.
31. Wu, F. C.; Lovorn, T.; MacDonald, A. H. *Physical Review Letters* **2017,** 118, (14), 147401.
32. Cao, B. X.; Li, T. S. *Journal of Physical Chemistry C* **2015,** 119, (2), 1247-1252.
33. Yeh, P. C.; Jin, W.; Zaki, N.; Kunstmann, J.; Chenet, D.; Arefe, G.; Sadowski, J. T.; Dadap, J. I.; Sutter, P.; Hone, J.; Osgood, R. M. *Nano Letters* **2016,** 16, (2), 953-959.
34. Yu, H. Y.; Liu, G. B.; Tang, J. J.; Xu, X. D.; Yao, W. *Science Advances* **2017,** 3, (11), e1701696.
35. Wu, F. C.; Lovorn, T.; Tutuc, E.; MacDonald, A. H. *Physical Review Letters* **2018,** 121, (2), 026402.
36. Wu, F. C.; Lovorn, T.; MacDonald, A. H. *Physical Review B* **2018,** 97, (3), 035306.
37. Wang, Y.; Wang, Z.; Yao, W.; Liu, G. B.; Yu, H. Y. *Physical Review B* **2017,** 95, (11), 115429.
38. Tong, W. Y.; Gong, S. J.; Wan, X. G.; Duan, C. G. *Nature Communications* **2016,** 7, 13612.
39. Shen, X. W.; Tong, W. Y.; Gong, S. J.; Duan, C. G. *2d Materials* **2018,** 5, (1), 011001.
40. Esters, M.; Hennig, R. G.; Johnson, D. C. *Physical Review B* **2017,** 96, (23), 235147.
41. Tong, W. Y.; Duan, C. G. *npj Quantum Materials* **2017,** 2, (1), 47.
42. Moon, P.; Koshino, M. *Physical Review B* **2013,** 87, (20), 205404.
43. Sato, K.; Saito, R.; Cong, C. X.; Yu, T.; Dresselhaus, M. S. *Physical Review B* **2012,** 86, (12), 125414.
44. Gong, S. J.; Gong, C.; Sun, Y. Y.; Tong, W. Y.; Duan, C. G.; Chu, J. H.; Zhang, X. *Proceedings of the National Academy of Sciences of the United States of America* **2018,** 115, (34), 8511-8516.
45. Gonzalez-Arraga, L. A.; Lado, J. L.; Guinea, F.; San-Jose, P. *Physical Review Letters* **2017,** 119, (10), 107201.
46. Sboychakov, A. O.; Rozhkov, A. V.; Rakhmanov, A. L.; Nori, F. *Physical Review Letters* **2018,** 120, (26), 266402.
47. Morell, E. S.; Vargas, P.; Chico, L.; Brey, L. *Physical Review B* **2011,** 84, (19), 195421.

48. Shallcross, S.; Sharma, S.; Pankratov, O. A. *Physical Review Letters* **2008,** 101, (5), 056803.
49. Koshino, M. *New Journal of Physics* **2015,** 17, 015014.
50. Jung, J.; Raoux, A.; Qiao, Z. H.; MacDonald, A. H. *Physical Review B* **2014,** 89, (20), 205414.
51. Kresse, G.; Furthmuller, J. *Computational Materials Science* **1996,** 6, (1), 15-50.
52. Perdew, J. P.; Burke, K.; Ernzerhof, M. *Physical Review Letters* **1996,** 77, (18), 3865-3868.
53. Grimme, S. *Journal of Computational Chemistry* **2006,** 27, (15), 1787-1799.